\documentstyle[11pt,newpasp,twoside,epsf]{article}
\begin{document}
\title{Optical studies of companions to millisecond pulsars}
\author{M. H. van Kerkwijk}
\affil{Department of Astronomy \& Astrophysics, University of Toronto,
  60 Saint George Street, Toronto, ON, M5S 3H8, Canada}
\author{C. G. Bassa}
\affil{Astronomical Institute, Utrecht University, PO Box 80000, 3508
  TA ~Utrecht, The Netherlands}
\author{B. A. Jacoby}
\affil{Department of Astronomy, California Institute of Technology, MC
  105-24, Pasadena, CA 91125, USA} 
\author{P. G. Jonker}
\affil{Center for Astrophysics, 60 Garden Street, Cambridge, MA 02138,
  USA} 

\begin{abstract}
Optical observations of the companions of pulsars can help determine
the properties of the binaries, as well as those of their components,
and give clues to the preceding evolution.  In this review, we first
describe the different classes of binary pulsars, and present a table
with a summary what is known about their optical counterparts.  Next,
we focus on the class of pulsars that have low-mass, helium-core white
dwarf companions.  We discuss attempts to determine the masses of both
components using optical spectroscopy, and compare the pulsar
spin-down ages with cooling ages of the white dwarfs.  We confirm that
for a given age, the lowest-mass white dwarfs are much hotter than the
more massive ones, consistent with recent evolutionary models,
although with one glaring exception.  We discuss the case of PSR
B0820+02, where the cooling age indicates a braking index less than 3,
and conclude by describing how cooling ages can be used to test
formation scenarios for PSR J1911$-$5958A, a pulsar binary in the
outskirts of NGC 6752.
\end{abstract}

\section{Binary pulsars and their evolutionary histories}

In Table~\ref{tab:optical}, we list all pulsars in binaries outside of
globular clusters.  One sees that their properties vary widely, but
one can identify different types on the basis of the spin and orbital
properties.  For instance, systems separate in clusters by inferred
companion mass and orbital period, as can be seen in
Fig.~\ref{fig:porbvsmass}.  Below, we briefly describe the different
groups and their evolutionary histories (for reviews, see Phinney \&
Kulkarni 1994; Stairs 2004).

\subsubsection{PSR+OB(e)} Pulsars with massive stellar
companions, which formed in binaries in which one star went supernova.
The pulsars are like young, isolated pulsars.  Probably, there are
many more PSR+OB(e) systems in which the pulsar is hidden by the
companion's stellar wind.  There should also be pulsars with
lower-mass companions, but as yet no secure identifications have been
made (a candidate is PSR B1820$-$11; Phinney \& Verbunt 1991).

\subsubsection{PSR+NS} Pulsars formed second in massive binaries, with
the first-formed neutron star as a companion.  For the one system
known, the first-born neutron star is a pulsar as well, but recycled
(see below). 

\subsubsection{PSR+CO/ONeMg-WD}  In binaries with two stars just below
the critical mass required to form a neutron star, the originally more
massive star will evolve first and leave a white dwarf.  In the
process, it may transfer sufficient amounts of matter to make the
originally lighter star massive enough to explode, leaving a newly
formed pulsar in an eccentric orbit around a massive white dwarf.

\subsubsection{Rec.-PSR+NS}  In binaries with a massive star and a
neutron-star companion, the star will eventually evolve.  Unless the
orbit is very wide, it will overflow its Roche lobe, and unstable mass
transfer to the neutron star will ensue, leading to a spiral in.  The
accretion of matter on the neutron star spins it up and, by mechanisms
unknown, reduces the magnetic field.  The resulting `recycled' pulsar
is left in a binary with another neutron star when the remaining
helium core is heavy enough to go supernova.  The orbit will be
eccentric.  These systems are also called `high-mass binary pulsars'
(HMBP).  For one system, the companion neutron star is a pulsar as
well.

\subsubsection{Rec.-PSR+CO-WD}  If, in the above scenario,
the helium core is not massive enough to explode, it will form a white
dwarf with a CO or ONeMg core and a helium envelope.  A massive white
dwarf can also be left if the companion of the pulsar evolved up to
the AGB before overflowing its Roche lobe.  The latter scenario may
lead to more accretion, and thus a neutron star spun up to faster
periods and with a more strongly reduced magnetic field.  The white
dwarf might still have a hydrogen envelope.  In either case, the orbit
will be circular.  These systems are also referred to as `intermediate
mass binary pulsars' (IMBP).

\subsubsection{Rec.-PSR+He-WD}  If the companion of a neutron star is a
low-mass star, the mass transfer that ensues when it evolves and
overfills its Roche lobe is stable.  A lot of mass is transferred,
leading to fast spin periods and low magnetic fields, as well as,
presumably, greatly increased mass.  If mass transfer started before
the helium flash, a helium-core white dwarf will be left, with a
hydrogen envelope.  If it started afterwards, a white dwarf with a CO
core will be formed, and the atmosphere may be either hydrogen or
helium.  In either case, one expects a circular orbit.  These systems
are also called `low-mass binary pulsars' (LMBP).

\subsubsection{Rec.-PSR+WD}  Recycled pulsars are also found with
companions with masses more similar to brown dwarfs.  These companions
likely originally had higher mass, but lost most of it in a long X-ray
binary phase.  The systems are also called `very low-mass binary
pulsars' (VLMBP) or, because of the strong irradiation and evaporation
observed, `black widow pulsars.'

\begin{figure}[t!]
\plotone{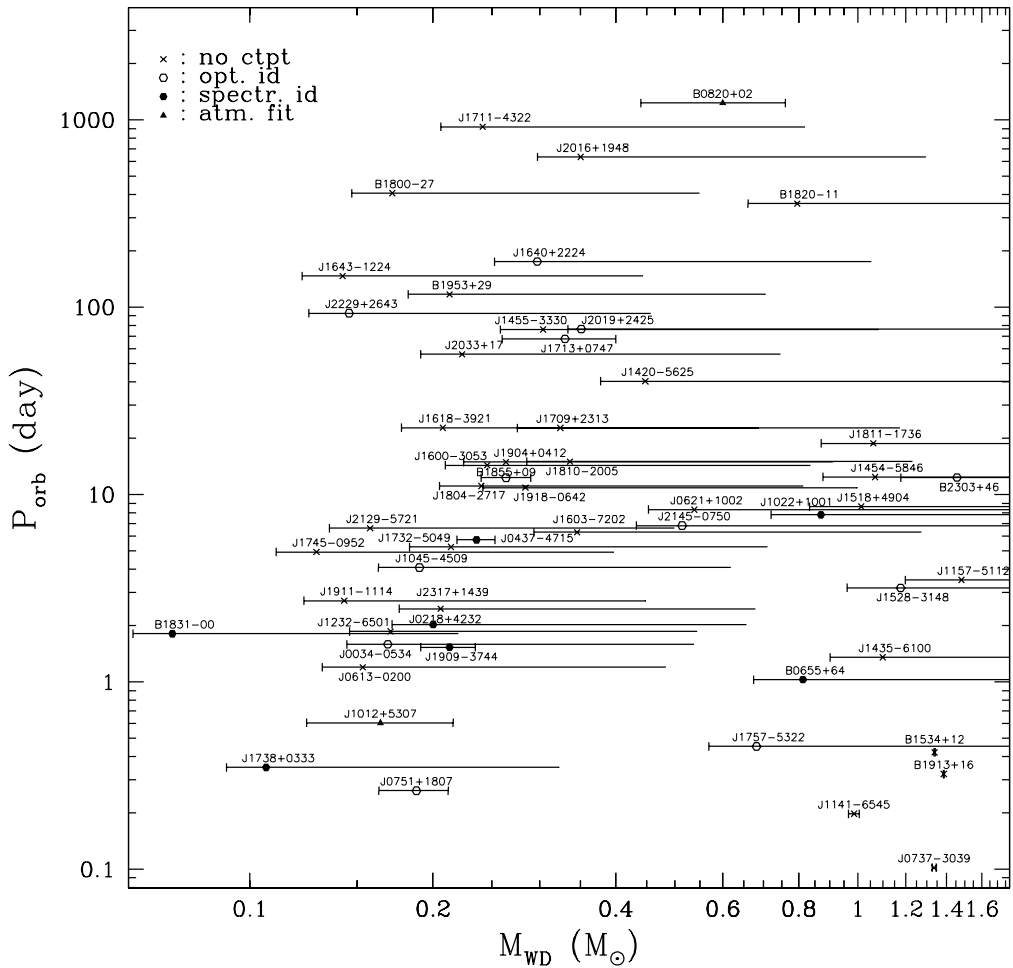}
\caption{Orbital period as a function of companion mass for all binary
  pulsars outside globular clusters with degenerate companions.
  Measured masses have 95\% confidence error bars.  For other systems,
  the masses are statistical, assuming a $1.4\,M_\odot$ pulsar mass
  and a $60^\circ$ inclination.  The horizontal bar indicates a range
  in inclination from $90^\circ$ (left handle) to $18^\circ$ (right
  side).  For each system, the marker indicates what is known
  about the optical counterpart (see legend and
  Table~\ref{tab:optical}).\label{fig:porbvsmass}}
\end{figure}

The various evolutionary scenarios lead to predictions for the
properties of the binary pulsars we observe.  In particular for the
recycled pulsars with low-mass white-dwarf companions, which evolved
via stable mass transfer, the predictions seem secure: there should be
a relation between companion mass and orbital period, one between
eccentricity and orbital period, and the neutron-star masses should
have increased (for a review, Phinney \& Kulkarni 1994).

Of these predictions, the second has been verified (Phinney \&
Kulkarni 1994), but the lack of accurate masses prevented stringent
tests of the other two.  With continuing high-precision timing,
however, the situation has changed.  For instance, Nice, Splaver, \&
Stairs (2004, these proceedings) have uncovered clear evidence that
the neutron stars are more massive than the typical 1.35\,$M_\odot$
inferred from double neutron-star binaries (Thorsett \& Chakrabarty
1999; these should have accreted little).

\begin{figure}[t!]
\plotone{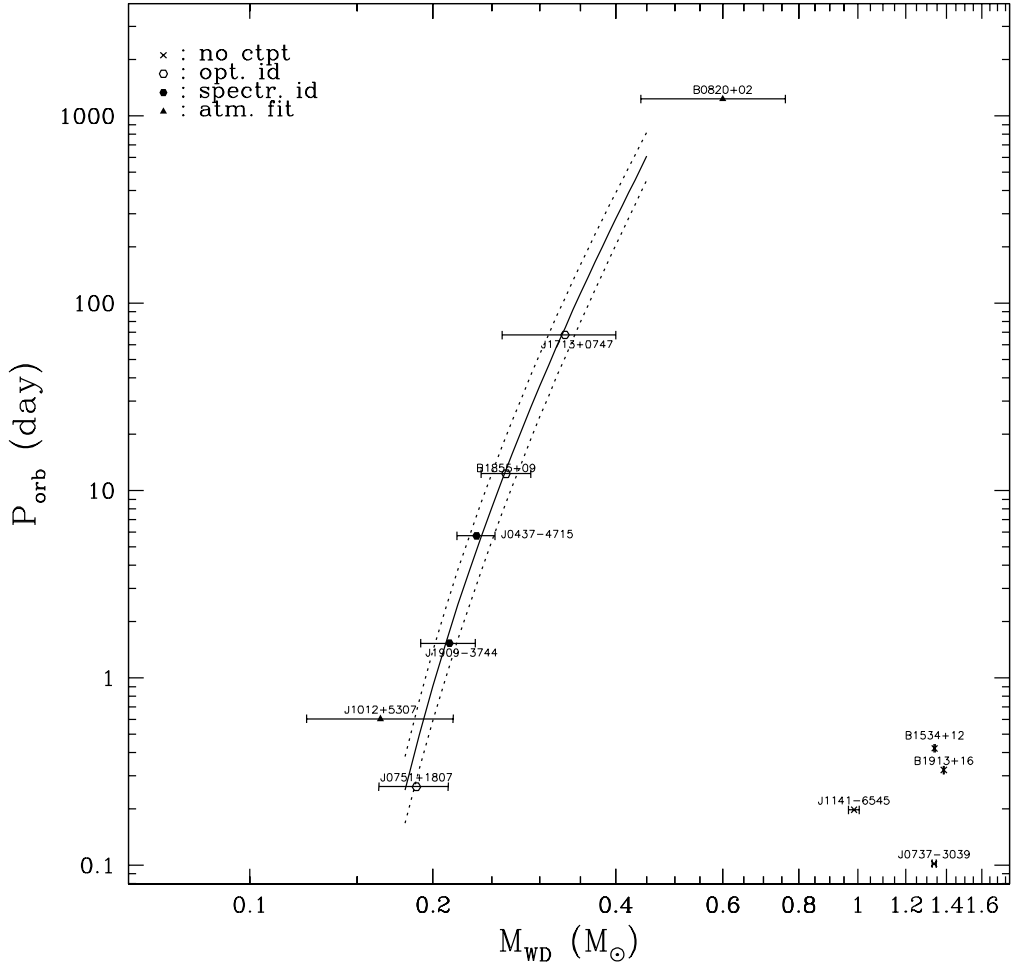}
\caption{Orbital period as a function of companion mass for all binary
  pulsars with measured masses (shown with 95\% confidence error
  bars).  Overdrawn are predictions from the evolutionary calculations
  of Tauris \& Savonije (1999).  The different lines are for different
  progenitor metallicities; there is some additional uncertainty
  related to the mixing-length parameter.\label{fig:porbmwd}}
\end{figure}

Timing measurements, as well as optical studies (see below), have also
yielded a number of accurate companion masses, which allow a much
improved test of the relation between companion mass and orbital
period.  In Fig.~\ref{fig:porbmwd}, we show these masses, with model
predictions of Tauris \& Savonije (1999) overdrawn.  The agreement is
impressive, especially when one considers that the models were
produced before the accurate masses became available.

\section{Optical counterparts}

In the table with all binary pulsars (Table~\ref{tab:optical}), we
also summarise what is known optically.  One sees that for the pulsars
with massive stellar companions, which should be bright, two out of
three are identified; the third (PSR J1740$-$3052) is highly obscured.
Also for the two `back widow' pulsars, the two strongly irradiated,
bloated, brown--dwarf mass companions have been identified.

\begin{table}[b!]
\caption[]{Optical properties of binary pulsars in the
  field.\label{tab:optical}} 
\begin{tabular}{@{}lr@{~~}r@{}r@{~~}rl@{}}
\hline\\[-2ex]
Name       &
\multicolumn{1}{c}{$P$}&
\multicolumn{1}{c}{DM}&
\multicolumn{1}{c}{$\log \tau_{\rm c}$}&
\multicolumn{1}{c}{$P_{\rm orb}$}&
Optical information\hfill [References]\\
           &
\multicolumn{1}{c}{(ms)}&
\multicolumn{1}{c}{\hbox to0pt{\hss(pc/cc)\hss}}&
\multicolumn{1}{c}{(yr)}&
\multicolumn{1}{c}{(d)}\\[0.3ex]
\hline\\[-2ex]
\multicolumn{4}{l}{\em PSR+OB(e)}\\  
J0045$-$7319 & 926.3& 105&  6.52&  51.17& B1V, 
                                          $V\!=\!16.19$ \hfill[kjb+94,bbs+95]\\
B1259$-$63   &  47.8& 147&  5.52&1236.72& B2e, 
                                          $V\!=\!10.05$ \hfill[jml+92,simbad]\\
J1740$-$3052 & 570.3& 741&  5.55& 231.03&                   \hfill[st++01]\\
\multicolumn{6}{l}{\em PSR+NS}\\     
J0737$-$3039\rlap{B}&2773.5&49&7.70&0.10\\
\multicolumn{4}{l}{\em PSR+CO/ONeMg-WD}\\     
J1141$-$6545 & 393.9& 116&  6.16&   0.20& $R\!>\!23.4$        \\
B2303$+$46   &1066.4&  62&  7.47&  12.34& $B\!=\!26.6$, 
                                          $B\!-\!R\!=\!0$   \hfill [vkk99]\\
\multicolumn{6}{l}{\em Rec.-PSR+NS (HMBP)}\\  
J0737$-$3039\rlap{A}&22.7&49&8.32&0.10\\
J1518$+$4904 &  40.9&  12& 10.37&   8.63& $B\!>\!24.5$, 
                                          $R\!>\!23$        \hfill [nst96]\\
B1534$+$12   &   3.8&  12&  7.39&   0.42\\
J1811$-$1736 & 104.2& 477&  8.96&  18.78\\
B1820$-$11$^{\rm a}$& 
               279.8& 429&  6.51& 357.76\\
B1913$+$16   &  59.0& 169&  8.03&   0.32\\
\multicolumn{6}{l}{\em Rec.-PSR+CO/ONeMg-WD (IMBP)}\\
J0621$+$1002 &  28.9&  37&  9.98&   8.32& $R\!>\!24$        \\
B0655$+$64   & 195.7&   9&  9.66&   1.03& DQ7, 
                                          $V\!=\!22.2$ \hfill [kul86,vkk95]\\
J1022$+$1001 &  16.5&  10&  9.78&   7.81& $V\!=\!23.1$, 
                                          $V\!-\!I\!=\!0.4$ \hfill [lfc96]\\
J1157$-$5112 &  43.6&  40&  9.68&   3.51& $R\!>\!23.7$:        \\
J1435$-$6100 &   9.3& 114&  9.78&   1.35& $R\!>\!23.1$         \\
J1454$-$5846 &  45.2& 116&  8.94&  12.42& $R\!>\!24.9$         \\
J1528$-$3146 &  60.8&  19&$\ldots$& 3.18& $R\!=\!23.9$:        \\
J1757$-$5322 &   8.9&  31&  9.73&   0.45& $R\!=\!24.6$:        \\
J2145$-$0750 &  16.1&   9&  9.93&   6.84& $V\!=\!23.7$, 
                                        $V\!-\!I\!=\!0.7$\hfill[lfc96]\\[0.2ex]
\hline\\[-2ex]
\end{tabular}

\noindent Note: Optical information without reference refers to
unpublished results of ourselves.  Colons indicate insecure
photometry.  For an overview of white-dwarf spectral types, see
Wesemael et al.\ (1993).  Briefly, `D' is for degenerate dwarf, `Q'
indicates the presence of carbon features in the spectrum, `A' the
presence of hydrogen, `C' the absence of any spectral features.  The
subtype $n$ is a measure of temperature, $T_{\rm eff}\simeq50400/n$.\\
$^{\rm a}$ B1820$-$11 may have a low-mass star as companion rather
than a neutron star (Phinney \& Verbunt 1991).
\end{table}

\begin{table}
\noindent\hbox to\hsize{\hskip2em Table 1 (Cont'd).\hskip1.5em Optical
  properties of binary pulsars in the field.\hfil} 
\begin{tabular}{@{}lr@{~~}r@{~~}r@{~~}rl@{}}
\hline\\[-2ex]
Name       &
\multicolumn{1}{c}{$P$}&
\multicolumn{1}{c}{DM}&
\multicolumn{1}{c}{$\log \tau_{\rm c}$}&
\multicolumn{1}{c}{$P_{\rm orb}$}&
Optical information~~~[References]\\
           &
\multicolumn{1}{c}{(ms)}&
\multicolumn{1}{c}{\hbox to0pt{\hss(pc/cc)\hss}}&
\multicolumn{1}{c}{(yr)}&
\multicolumn{1}{c}{(d)}\\[0.3ex]
\hline\\[-2ex]
\multicolumn{4}{l}{\em Rec.-PSR+He-WD (LMBP)}\\
J0034$-$0534 &   1.9&  14&  9.78&   1.59& $I\!=\!24.8$, 
                                          $V\!-\!I\!>\!2.0$ \hfill [lfc96]\\
J0218$+$4232 &   2.3&  61&  8.68&   2.03& DA6, 
                                          $V\!=\!24.2$      \hfill [bvkk03]\\
J0437$-$4715 &   5.8&   3&  9.20&   5.74& DC12, 
                                          $V\!=\!20.8$      \hfill [dbdv93]\\
J0613$-$0200 &   3.1&  39&  9.71&   1.20& brightish star nearby$^{\rm b}$\\
J0751$+$1807 &   3.5&  30&  9.85&   0.26& $R\!=\!25.1$, 
                                          $R\!-\!I\!=\!0.9$ \hfill [bvkk04]\\
B0820$+$02$^{\rm a}$&
               864.9&  24&  8.12&1232.40& DA3, 
                                          $V\!=\!22.8$      \hfill [kr00]\\
J1012$+$5307 &   5.3&   9&  9.69&   0.60& DA6, 
                                          $V=19.6$     \hfill [llfn95,vkbk96]\\
J1045$-$4509 &   7.5&  58&  9.83&   4.08& $R\!\sim\!24$:    \\
J1232$-$6501 &  88.3& 239&  9.24&   1.86& $R\!>\!24$        \\
J1420$-$5625 &  34.1&  65&  9.90&  40.29& crowded field     \\
J1455$-$3330 &   8.0&  14&  9.72&  76.17& $R\!>\!24$        \\
J1600$-$3053 &   3.6&  52&  9.78&  14.35& $R\!>\!24$        \\
J1603$-$7202\rlap{$^{\rm a}$}&
                14.8&  38& 10.18&   6.31& $R\!>\!24$ (v.faint ctpt?)\\
J1618$-$3921 &  12.0& 118&  9.55&  22.80& $R\!>\!24$        \\
J1640$+$2224 &   3.2&  18& 10.25& 175.46& $V\!=\!26.0$, 
                                          $V\!-\!I\!=\!1.4$ \hfill [lfc96]\\
J1643$-$1224 &   4.6&  62&  9.60& 147.02& $R\!\sim\!23$::   \\
J1709$+$2313 &   4.6&  25& 10.31&  22.71& $R\!>\!24$        \\
J1711$-$4322 & 102.6& 192&  6.89& 920.2 & pos.\ too unc.\   \\
J1713$+$0747 &   4.6&  16&  9.93&  67.83& $V\!=\!26.0$, 
                                          $V\!-\!I\!=\!1.9$ \hfill [lfc96]\\
J1732$-$5049 &   5.3&  57&  9.79&   5.26& $R\!>\!24$        \\
J1738$+$0333 &   5.8&  34&  9.61&   0.35& DA6, $V\!\sim\!21$\\
J1745$-$0952 &  19.4&  64&  9.51&   4.94& $R\!>\!24$        \\
B1800$-$27   & 334.4& 166&  8.49& 406.78& crowded field     \\
J1804$-$2717 &   9.3&  25&  9.56&  11.13& $R\!>\!24$        \\
J1810$-$2005 &  32.8& 240&  9.54&  15.01& $R\!>\!24$        \\
B1831$-$00\rlap{$^{\rm a}$}& 
               521.0&  89&  8.89&   1.81& $R\!=\!22.0$, 
                                          $R\!-\!K\!=\!2.3$ \\ 
B1855$+$09   &   5.4&  13&  9.68&  12.33& $V\!=\!25.9$, 
                                          $V\!-\!I\!=\!1.7$ \hfill [vkbkk00]\\
J1904$+$0412 &  71.1& 186& 10.01&  14.93& $R\!>\!24$        \\
J1909$-$3744 &   2.9&  10&  9.52&   1.53& DA6, $V\!\sim\!21$\hfill [jbvk+03]\\
J1911$-$1114 &   3.6&  31&  9.61&   2.72& $R\!>\!24$ (v.\ faint ctpt?)\\
J1918$-$0642 &   7.6&  27&  9.70&  10.91& $R\!>\!24$        \\
B1953$+$29   &   6.1& 105&  9.51& 117.35& brightish star nearby$^{\rm b}$\\
J2016$+$1948 &  64.9&  34&$\ldots$&635.04\\
J2019$+$2425 &   3.9&  17&  9.95&  76.51& $I\!=\!25.0$, 
                                          $V\!-\!I\!>\!1.1$ \hfill [lfc96]\\
J2033$+$17   &   5.9&  25&  9.93&  56.31& pos.\ too unc.\   \\
J2129$-$5721 &   3.7&  32&  9.45&   6.63& $R\!>\!24$        \\
J2229$+$2643 &   3.0&  23& 10.51&  93.02& $R\!\sim\!25$:    \\
J2317$+$1439 &   3.4&  22& 10.35&   2.46& $R\!>\!24$        \\
\multicolumn{4}{l}{\em Rec.-PSR+BD (VLMBP)}\\
B1957$+$20   &   1.6&  29&  9.18&   0.38& $R\!=\!19.4\ldots>\!24$
                                                       \hfill[cvpr95,fbb95]\\
J2051$-$0827 &   4.5&  21&  9.75&   0.10& $R\!=\!22.5\ldots25.5$
                                                       \hfill[svkbk01]\\[0.2ex]
\hline\\[-2ex]
\end{tabular}

\noindent$^{\rm a}$ B0820+02 has a CO-WD companion.  J1603$-$7202 may
be an IMBP, given its high mass function.  B1831$-$00 is
hardly recycled; it may have formed differently.\\ 
$^{\rm b}$ These stars are likely unassociated, but prevent deep
searches.
\end{table}

For the pulsars likely to have white-dwarf companions, 22 out of 49
have been identified optically (see also Fig.~\ref{fig:porbvsmass}).
Of these, 7 have spectral types and a further 9 have some colour
information.  As expected from evolutionary considerations, the
spectral type of the one more massive companion indicates a helium
atmosphere, while a hydrogen atmosphere is found for companions
inferred to be low-mass white dwarfs (e.g., Van Kerkwijk \& Kulkarni
1995).

For most of the 27 unidentified pulsar, white dwarf binaries, there
are upper limits of roughly 24th magnitude.  These are predominantly
the result of a systematic campaign to find all objects bright enough
to do spectroscopy, which, as will become clear below, allows one to
obtain the most interesting results.

\section{Radial velocities and masses}

For the counterparts bright enough for spectroscopy, one can model the
spectrum, and determine a precise temperature and surface gravity for
the white dwarf.  Combined with white-dwarf mass-radius relations from
cooling models, these yield the white-dwarf mass, which can be used to
verify the predictions from binary evolution.  If the orbit is short
enough, one can also measure radial velocities and determine the
radial-velocity amplitude.  Combined with the precise radial-velocity
amplitude of the pulsar (derived from timing), this yields the mass
ratio and thus, with the mass of the white dwarf, the neutron-star
mass.

For the first relatively bright counterpart discovered, that of PSR
J0437$-$4715, the result was disappointing: the spectrum was
featureless (Danziger et al.\ 1993).  The temperature of
$\sim\!4000\,$K is too low for any features to appear.

For the even brighter counterpart of PSR J1012+5307, however, strong
hydrogen lines were present in the spectrum, and a model-atmosphere
and radial-velocity analysis were done by two groups (Van Kerkwijk,
Bergeron, \& Kulkarni 1996; Callanan, Garnavich, \& Koester 1998).
Unfortunately, the results were less constraining than hoped.  First,
the radial-velocity amplitudes found by the two teams were different.
It turned out this was related to a reduction error by Van Kerkwijk et
al.\ (1996); a re-reduction of the original data, complemented with
more recent results, yields a radial-velocity amplitude of
$199\pm10{\rm\,km\,s^{-1}}$, consistent with the
$218\pm10{\rm\,km\,s^{-1}}$ found by Callanan et al.\ (1998).  This
result should still improve, once small remaining systematic effects
have been taken into account.  From the two estimates, the current
best estimate of the mass ratio is $M_{\rm NS}/M_{\rm WD}=10.0\pm0.7$.

A second discrepancy between the two studies was the value of the
surface gravity inferred: Van Kerkwijk et al.\ (1996) found $\log
g=6.75\pm0.07$, while Callanan et al.\ (1998) inferred $\log
g=6.34\pm0.20$.  For this difference, the underlying cause turned out
to be two different sets of model atmospheres used (by P.\ Bergeron
and D.\ Koester, respectively).  When the spectra of Van Kerkwijk et
al.\ were fitted using the Koester models, a surface gravity
consistent with that of Callanan et al.\ is found (D.\ Koester, pers.\
comm.).

Finally, it was found that for the very low masses involved, the
mass-radius relation was less securely known than expected.
Coincidentally, the slightly different approaches taken by the two
teams compensated the differences in surface gravity, leading to the
same final white dwarf mass, $0.16\pm0.02\,M_\odot$.  

The above mass and mass ratio correspond to a neutron star mass of
$1.6\pm0.2\,M_\odot$, where the uncertainty is dominated by the
uncertainty in the white-dwarf mass.  If instead we use the
white-dwarf mass expected from the models of Tauris \& Savonije
(1998), $M_{\rm WD}=0.193\pm0.007$ (see Fig.~\ref{fig:porbmwd}), we
infer a neutron star mass of $1.9\pm0.2\,M_\odot$.  This is a large
mass, but similar to is found for PSR J0751+1807 by Nice et al.\
(2004, these proceedings).  Clearly, it will be worthwile to try to
determine the white-dwarf mass more accurately.

The above procedure was also tried on another short-period pulsar
binary, PSR J0218+4232.  This system is substantially fainter
($V=24.2$; Table~\ref{tab:optical}), and, unfortunately, it turned out
to be beyond the capabilities of the then-available instrumentation:
while an accurate temperature could be determined, no useful
constraints could be derived on the surface gravity and
radial-velocity amplitude (Bassa, Van Kerkwijk, \& Kulkarni 2003a).

Obviously, the current large uncertainties are somewhat discouraging.
Fortunately, on all fronts improvements are being made.  First, two
more bright counterparts to recycled pulsars have been discovered, to
PSR J1909$-$3744 and J1738+0333 (see Table~\ref{tab:optical}).  Both
have spectra similar to J1012+5307, and thus similar temperatures and
surface gravities.  Of the two, PSR J1909$-$3744 is particularly
interesting, as the orbit is sufficiently edge-on to allow for an
accurate determination of the white-dwarf mass from Shapiro delay
(Jacoby et al.\ 2003).  Combined with much improved white-dwarf
mass-radius relations (from detailed cooling models; see below), this
implies we will know the surface gravity of the white dwarf, which we
can use to calibrate the model-atmosphere analysis.

The second improvement is that more blue-sensitive and more stable
spectrographs are now available on large telescopes, and hence more
precise radial-velocity curves can be measured.  It should be possible
to obtain mass ratios to better than $\sim\!5$\% accuracy.  Observing
campaigns of both new pulsar binaries are underway.

We conclude by mentioning one last system, PSR B0820+02, for which it
has been possible to derive an accurate companion mass,
$M=0.60\pm0.08$ (Koester \& Reimers 2000).  This mass is similar to
the masses found for isolated white dwarfs, for which the white-dwarf
model atmospheres and mass-radius relations are well understood.
Hence, it should be reliable.  Unfortunately, the system is unsuitable
for determining the neutron star mass, as the orbit is too wide to
determine an accurate radial-velocity curve.  However, as we will see
below, the accurate white-dwarf mass and temperature allow for
interesting constraints on the cooling age.

\section{Cooling versus spin-down age}

After the last bit of mass is transferred from the progenitor of the
white dwarf to the neutron star, two independent clocks start running.
The first clock is the millisecond pulsar, which will turn on and
start slowing down.  Assuming a spin-down torque $\propto\nu^n$, the
pulsar age is given by
\begin{equation}
t_{\rm psr} = \frac{P}{(n-1)\dot{P}}
        \left[1-\left(\frac{P_0}{P}\right)^{n-1}\right],
\label{eq:tpsr}
\end{equation}
where $P\equiv1/\nu$ is the current spin period, $\dot{P}$ is its rate
of change, $P_0$ is the period when the pulsar began spinning down
following cessation of mass transfer, and
$n=\nu\ddot{\nu}/\dot{\nu}^2$ is the ``braking index,'' equal to 3
under the assumption of magnetic dipole radiation.  For $n=3$ and
$P_0\ll{}P$, $t_{\rm psr}\simeq\tau_{\rm c}\equiv{}P/2\dot{P}$, where
$\tau_{\rm c}$ is the pulsar ``characteristic age.''

The second clock is the white dwarf.  After the remaining envelope has
been burned off, the white dwarf can only radiate its internal heat,
making it cool down as time goes by.  In principle, it is fairly easy
to estimate the cooling, as the thermal structure of the white dwarf
is simple.  The heat is stored in the non-degenerate ions in the
white dwarf interior, which is kept nearly isothermal due to the
efficient heat conduction by degenerate electrons.  The cooling rate
is determined by the much less efficient radiative heat transport near
the white dwarf's atmosphere.

In practice, there are complications.  Apart from difficulties in
modelling the radiative opacities and dealing with convection, there
are two additional physical processes that play a role.  First, at low
temperatures, the ion gas in the core starts to crystallise.  The
latent heat released temporarily keeps the white dwarf warmer, but
once gone, allows for much more rapid cooling.  This effect is
particularly important for more massive, carbon-oxygen white dwarfs.
Second, for white dwarfs with relatively thick residual hydrogen
envelopes, the pressure at the bottom can be sufficiently high for
significant pycno-nuclear fusion, keeping the white dwarf warm longer.
We will return to this below.

The possible use of comparing pulsar and white-dwarf ages was realized
immediately upon the first detection of optical emission from
white-dwarf counterparts (Kulkarni 1986).  For the white dwarf
accompanying PSR B0655+64, a relatively low temperature,
$\sim\!7000\,$K, was inferred, which implied an age of $\sim\!2\,$Gyr.
This meant that pulsar magnetic fields could not decay completely on
this timescale, as had been common wisdom at the time.

The first systematic comparison between spin-down and cooling ages was
done by Hansen \& Phinney (1998a,b).  Since they had to rely on
estimated masses, their results were uncertain, but two clear
discrepancies stood out: for PSR J1012+5307, the inferred cooling age
was far shorter than the characteristic age, while for PSR B0820+02,
the cooling age was significantly longer.  We discuss both
discrepancies in turn.

\subsection{PSR J1012+5307 and other short-period binaries}

Already in their discovery paper of PSR J1012+5307's optical
counterpart, Lorimer et al.\ (1995) noted that the white dwarf was
much hotter than expected given the pulsar's characteristic age.  They
suggested that the problem might lie in the pulsar age: if the initial
period to which the pulsar was spun up was similar to the current one
(i.e., $P_0\simeq P$ instead of $P_0\ll P$ in Eq.~1), then the pulsar
age could be equal to a short cooling age.  Alberts et al.\ (1996),
however, suggested the white-dwarf age had been underestimated: given
the low mass, the hydrogen layer on the white dwarf could be quite
thick, and residual hydrogen burning could keep the white dwarf hot.
Indeed, this effect had already been found by Webbink (1975), in a
general study of the evolution of helium white dwarfs in close
binaries.

A flurry of modelling followed, confirming the likely presence of a
thick hydrogen layer (Driebe et al.\ 1998, 1999), and adding
complications, such as the duration of the semi-detached phase in
which the companion is becoming a white dwarf (Sarna, Antipova, \&
Muslimov 1998).  At first, the new models seemed to suggest that most
low-mass white-dwarf companions should be fairly bright, but this was
disproven by the discovery of a very faint counterpart to PSR B1855+09
(Van Kerkwijk et al.\ 2000).  This led to more careful considerations
of the effects of shell flashes during the formation (e.g.,
Sch\"onberner, Driebe, \& Bl\"ocker 2000), and to the effects of
element diffusion on these flashes (Sarna, Ergma, \& Ger{\v s}kevit{\v
s}-Antipova 2000; Althaus, Serenelli, \& Benvenuto 2001a,b).  

The current consensus appears to be that below a certain critical
mass, the hydrogen layer should be thick and white dwarfs should be
relatively hot, $\sim\!10^4\,$K even when several Gyr old.  Above the
critical mass, shell flashes occur, and the hydrogen layer will be too
thin to sustain significant residual fusion; after several Gyr, those
more massive white dwarfs will have cooled to below 5000\,K.  The
precise value of the critical mass is not known, but, since the shell
flashes result from CNO burning, almost certainly depends on
metallicity (Sarna et al.\ 2000; Serenelli et al.\ 2002).

\begin{table}[t!]
\caption[]{Hydrogen layer properties for helium WD
  companions.\label{tab:hlayer}}  
\begin{tabular}{@{}lrrrrclr@{}}
\hline\\[-2ex]
Name       &
\multicolumn{1}{c}{$d_{\rm TC}{}^{\rm a}$}&
\multicolumn{1}{c}{$d_{\rm CL}{}^{\rm a}$}&
\multicolumn{1}{c}{$\log \tau_{\rm c}$}&
\multicolumn{1}{c}{$P_{\rm orb}$}&
\multicolumn{1}{c}{$M_{\rm WD}{}^{\rm b}$}&
\multicolumn{1}{c}{$T_{\rm eff}{}^{\rm c}$}&
H layer\\
           &
\multicolumn{1}{c}{(kpc)}&
\multicolumn{1}{c}{(kpc)}&
\multicolumn{1}{c}{(yr)}&
\multicolumn{1}{c}{(d)}&
\multicolumn{1}{c}{$(M_\odot)$}&
\multicolumn{1}{c}{(kK)}\\[0.3ex]
\hline\\[-2ex]
J0751$+$1807 &  2.0&1.1&     9.85&   0.26&~0.16--0.21~ &~~3   & thin\\
J1738$+$0333 &  1.9&1.4&     9.61&   0.35&           &~~9   &thick\\
J1012$+$5307 &  0.5&0.4&     9.69&   0.60&0.12--0.20 &~~8.5 &thick\\
J0613$-$0200 &  2.2&1.7&     9.71&   1.20\\
J1909$-$3744 &\multicolumn{2}{c}{$1.1(+0.3/-0.2)$}& 
                             9.52&   1.53&0.19--0.22 &~~8   &thick\\
J0034$-$0534 &  1.0&0.5&     9.78&   1.59&           &~~\llap{$<$\,}4& thin\\
J1232$-$6501 &10\phantom{.0}&6.2& 9.24&  1.86\\
J0218$+$4232 &  5.9&2.6&     8.68&   2.03&           &~~8   & thin\\
J2317$+$1439 &  1.9&0.8&    10.35&   2.46&           &~~\llap{$<$\,}4:& thin\\
J1911$-$1114 &  1.6&1.2&     9.61&   2.72&           &~~\llap{$<$\,}5:& thin\\
J0437$-$4715 &\multicolumn{2}{c}{$0.139\pm0.003$}&
                             9.20&   5.74&0.20--0.27 &~~4   & thin\\
B1855$+$09   &\multicolumn{2}{c}{$0.9(+0.4/-0.2)$}& 
                             9.68&  12.33&0.24--0.40 &~~5   & thin\\
J1713$+$0747 &\multicolumn{2}{c}{$1.1(+0.5/-0.3)$}& 
                             9.93&  67.83&           &~~3.5 & thin\\
B0820$+$02$^{\rm d}$&\multicolumn{2}{c}{$1.86\pm0.13$}&
                             8.12&1232.40&0.44--0.76 &~~\llap{1}5& thin\\
\hline\\[-2ex]
\end{tabular}

\noindent$^{\rm a}$ Distances are inferred from the dispersion measure
using the Taylor \& Cordes (1993) and the Cordes \& Lazio (2002) models
of the Galactic electron distribution.  Distances with uncertainties
(1$\sigma$) are from parallax measurements (J1909: Jacobi et al., in
prep.; J0437: [vsbb+01]; B1855: [ktr94]; J1713: [cfw94]) and from
modelling the white-dwarf spectrum (B0820: [kr00]).\\
$^{\rm b}$ Masses are from Shapiro delay (J0751, B1855, J1713:
[nss04]; J1909: Jacobi et al., in prep.; J0437: [vsbb+01]), and from
modelling the white-dwarf spectrum (J1012: [vkbk96,cgk98]; B0820:
[kr00]).\\ 
$^{\rm c}$ Temperatures are from colours or spectra, except for
J2317+1439 and J1911$-$1114, for which the limits were derived from
the magnitude limit, combined with the Cordes \& Lazio distance.
These are marked with a colon to indicate they are uncertain.\\
$^{\rm d}$ B0820+02 has a CO WD companion, but formed like a LMBP.
\end{table}

Until recently, there was not much supporting observational evidence.
Indeed, except for PSR J1012+5307, all identified white dwarf
counterparts were rather cool, implying thin hydrogen envelopes.
Could it be that thick envelopes did not occur at all, and that PSR
J1012+5307 was re-born with $P_0\simeq P$ after all (Bassa et al.\
2003a)?

Additional evidence for the presence of thick hydrogen envelopes was
uncovered with the identification of rather bright and hot white-dwarf
counterparts to two new pulsars with large characteristic ages (PSR
J1909$-$3744 and PSR J1738+0333; see Table~\ref{tab:optical}).  With
those, it has become possible to try to determine empirically the
critical mass below which hydrogen layers will be thick.
Unfortunately, we do not have accurate masses for most white-dwarf
companions.  However, we can use the orbital period as a proxy, as
this should correlate fairly tightly with the companion mass (see
Fig.~\ref{fig:porbmwd} and Section 1).

In Table~\ref{tab:hlayer}, we list the properties for those low-mass
binary pulsars that either have periods less than 3 days or have
measured companion masses.  The final column lists whether a thick or
a thin hydrogen envelope is required to understand the observed
temperature (assuming the white-dwarf cooling age is similar to the
characteristic age).  One sees that for all systems with orbital
periods in excess of 1.55\,d, thin hydrogen envelopes are inferred,
while below that period they are likely thick (with one exception; see
below).  For the system just below the critical period, PSR
J1909$-$3744, we have a mass estimate from Shapiro delay.  The
preliminary results from Jacoby et al.\ (2004, in prep.), is
$0.203\pm0.006\,M_\odot$.  This is slightly more massive than found
from the models of, e.g., Althaus et al.\ (2001b), which suggest that
$0.196\,M_\odot$ white dwarfs should still have thin envelopes.
However, given the uncertainties (e.g., in metallicity; Serenelli et
al.\ 2002), the agreement seems satisfactory.

While the above paints a consistent picture, there is one exception:
for PSR J0751$-$1807, a low temperature is measured (Bassa, Van
Kerkwijk, \& Kulkarni 2004).  A low temperature had already been
indicated by earlier limits of Lundgren et al.\ (1996a), and Ergma,
Sarna, \& Ger{\v s}kevit{\v s}-Antipova (2001) suggested the hydrogen
envelope might have been reduced shortly after the mass transfer
ceased, due to irradiation by the pulsar (which would be particularly
effective in such a close binary).  Intriguingly, however, our
photometry seems to indicate that the white dwarf has no hydrogen
whatsoever, but instead a helium atmosphere.

\subsection{PSR B0820+02 and its surprisingly cool white dwarf}

The second discrepancy identified by Hansen \& Phinney (1998b) was
that the cooling age for PSR B0820+02 was significantly longer than
the characteristic age.  This was confirmed by Koester \& Reimers
(2000), who analysed spectra of the optical companion and inferred
$T_{\rm eff}=15000\pm800\,$K and $\log g=7.98\pm0.13$.  Combined with
cooling models, this yields a mass of $0.60\pm0.08\,M_\odot$.  This
mass implies a CO core, and is typical for an isolated white dwarf,
not unexpected given the very wide orbit, and the fact that the pulsar
appears to have been recycled only very mildly ($P=0.9\,$s;
$B\simeq3\times10^{11}\,$G).

For a mass of $0.6\,M_\odot$, the implied cooling age is
$221\pm11\,$Myr, which is significantly above the pulsar's
characteristic age, $\tau_{\rm c}=130\,$Myr.  Koester \& Reimers
(2000) argue that the white-dwarf cooling models for these masses are
secure, and hence that the pulsar age must be larger than the
characteristic age.  From Eq.~1, one sees that the only way to do that
is to decrease the breaking index $n$; consistency with the cooling
age would then require $n=2.2$.

If true, the above implies that pulsar spin-down is different from
magnetic dipole braking not just for young pulsars (e.g., Lyne 1996),
but also for older ones.  As yet, however, the conclusion is less firm
than it might appear: the uncertainty in the age listed by Koester \&
Reimers (2000) does not seem to include the uncertainty in the mass of
the object.  For a less massive white dwarf, the cooling age would be
reduced (e.g., 150\,Myr for a $0.5\,M_\odot$ white dwarf; see also
Sch\"onberner et al.\ 2000).  This can be verified with better
spectra.

\subsection{Application to PSR J1911$-$5958A in NGC 6752}

With the cooling models in quantitative agreement with the
observations, it has become possible to use them.  One particularly
interesting case is PSR J1911$-$5958A in the globular cluster NGC
6752.  This binary is puzzling as it is very far, $\sim\!3.3$
half-mass radii outside the core (D'Amico et al.\ 2002; also Possenti
et al.\ 2004, these proceedings), and it is unclear how it could have
gotten there.  Colpi et al.\ (2002) investigated different
possibilities, and found it was difficult to produce the system either
from a primordial binary or by a scattering or exchange event.
Instead, they suggested that the binary may have been scattered by a
binary composed of two fairly massive black holes.  In this scenario,
the binary would already have formed before the scattering event.

An age estimate could be used to distinguish between the various
possibilities.  For a primordial origin or an older binary being
scattered by a binary black hole, the white dwarf would most likely be
rather old.  If the white dwarf was formed during or shortly after an
exchange, however, it could not be much older than the
$\sim\!0.7\,$Gyr the binary can be expected to stay in the outskirts
if it is currently on a highly eccentric orbit in the cluster (Colpi
et al.\ 2002).

It turned out that archival ESO Wide Field Imager and {\em HST} WFPC2
images were available of the field, and those allowed us to identify
the white-dwarf companion of PSR J1911$-$5958A (Bassa et al.\ 2003b;
the source was also identified in new VLT observations by Ferraro et
al.\ 2003).  It is relatively bright and hot, with a best-fit
temperature of $\sim\!11000\,$K.  As the distance is known, we have a
measure of the radius; the implied mass is low, about $0.2\,M_\odot$,
as expected from the pulsar mass function.  

The high temperature implies that the white dwarf is rather young,
$\sim\!1\,$Gyr, even if it has a thick hydrogen layer (at the low
cluster metallicity, the critical mass is about $0.22\,M_\odot$;
Serenelli et al.\ 2002).  This suggests that the system was formed in
a simple exchange interaction with another star or binary in the
core.  

Independent of its formation, the system is interesting simply because
it is sufficiently bright, $V\simeq22$, to allow spectroscopy, and
measure the white dwarf and neutron star masses using the procedure
outlined in Section 3.  An advantage over systems in the field is that
the distance is known, which means one can measure the white-dwarf
radius using the flux and temperature, and thus have a second
constraint on the white-dwarf mass.

\acknowledgements We thank David Nice for sharing his timing results
in advance of publication, and the conference organisers for granting
us additional time to write up this review.

\end{document}